\documentclass[a4paper]{jpconf}
\usepackage{amsfonts}
\usepackage{amsmath}
\usepackage{bm}
\usepackage{bbm}
\usepackage{xcolor}
\usepackage{subfig}
\usepackage{graphicx}
\usepackage[utf8]{inputenc}
\usepackage{caption}

\newcommand{\la}{\langle}
\newcommand{\ra}{\rangle}
\newcommand{\ua}{\uparrow}
\newcommand{\da}{\downarrow}

\newcommand{\co}[2]{\hat{#1}^\dagger_{#2}}
\newcommand{\cs}[2]{\hat{#1}^\dagger_{#2\sigma}}

\newcommand{\ao}[2]{\hat{#1}_{#2}}
\newcommand{\as}[2]{\hat{#1}_{#2\sigma}}
\newcommand{\au}[2]{\hat{#1}_{#2\ua}}
\renewcommand{\ad}[2]{\hat{#1}_{#2\da}}

\newcommand*{\Scale}[2][4]{\scalebox{#1}{\ensuremath{#2}}}%
\newcommand{\Eq}[1]{Eq. (\ref{#1})}
\begin{document}
\title{Computing Exact Self-Energies with Polynomial Expansion}

\author{M. Hyrk\"as, D. Karlsson and R. van Leeuwen}

\address{Nanoscience Center, Department of Physics, FIN-40014 University of Jyv\"askyl\"a, Finland}

\ead{markku.hyrkas@jyu.fi}

\begin{abstract}
We give details on how to calculate spectral functions and Green's functions for finite systems using the 
Chebyshev polynomial expansion method. We apply the method to a finite Anderson impurity system, and furthermore give details on how to exploit its symmetry to transform the system into smaller orthogonal subsystems. We also consider systems connected to infinite leads, which we study by approximating the unknown self-energy with an exact self-energy for a finite system. As our test case, we consider the single-impurity Anderson model, where we find that we can capture some aspects of Kondo physics.
\end{abstract}

\section{Introduction}

Transport through strongly correlated system, where a perturbative treatment of interactions fails, is a difficult yet important field of study with implications for molecular electronics, impurity effects in materials etc. It continues to be a subject of intense research, and several methods to deal with such systems have been developed \cite{Heidrich-Meisner2009, Jacob2010, Arrigoni2013}. If the interactions can be assumed to be localized in a small region, exact diagonalization methods can be brought to bear on the problem \cite{Ishida2012}. A successful strategy, the Embedded Cluster Approximation\cite{Ferrari1999, Heidrich-Meisner2009}, has been to split from the infinite system a finite part that can be treated with accurate methods, and then connected to the rest of the system perturbatively on the Green's function level.

In this work we will follow a similar strategy, but we instead take the self-energy of the finite system as a starting point for approximating the infinite system \cite{Hubbard1963, Gros1993}. By making use of the Chebyshev polynomial method, we can treat larger finite systems exactly, and thus improve the quality of our approximate self-energies.

The system under study consists of a finite interacting cluster connected (embedded) to several infinite non-interacting leads ($\alpha$), described by the Hamiltonian
\begin{equation}
\hat{H} = \hat{H}_{f,0} + \hat{H}_{f,int} + \sum_\alpha \left( \hat{H}^\alpha_0 + \hat{H}^{\alpha}_{em}\right),
\end{equation}
where $\hat{H}_{f,0} + \hat{H}_{f,int}$ describes the finite system (with $\hat{H}_{f,int}$ containing the interactions), $\hat{H}^\alpha_0$ the non-interacting lead $\alpha$, and $\hat{H}^{\alpha}_{em}$ the lead-cluster connection.

We want to obtain the retarded Green's function $G$ in the cluster in the presence of the leads, to gain access to observables like single-particle densities, conductances and spectral functions. Our starting point is the Dyson equation for $G$ in equilibrium
\begin{equation} \label{eq:dyson}
\bm{G}(\omega) = \bm{G}_{f,0}(\omega) + \bm{G}_{f,0}(\omega) \Big[ \bm{\Sigma}(\omega) + \sum_\alpha \bm{\Sigma}^\alpha_{em}(\omega) \Big] \bm{G}(\omega).
\end{equation}
The terms in this equation are matrices, their elements corresponding to particle/hole propagation between any two sites of the cluster. $\bm{G}_{f,0}(\omega)$ is the non-interacting non-embedded Green's function for the cluster, and $\bm{\Sigma}(\omega)$ and $\bm{\Sigma}_{em}(\omega)$ are self-energies describing the interaction and the coupling to the leads (embedding), respectively \cite{Stefanucci2013}. 
The precise form of $\bm{\Sigma}^\alpha_{em}(\omega)$, which we assume to be known, depends only on the structure of the leads, and can be calculated separately.

Our main focus is the interaction self-energy $\bm{\Sigma}(\omega)$, which describes the correlations in the system. It is affected by the presence of the infinite leads, and therefore has no simple expression and has to be approximated. A simple way forward is to replace it with the interaction self-energy of the disconnected cluster, for which we have
\begin{equation}
\bm{G}_f(\omega) = \bm{G}_{f,0}(\omega) + \bm{G}_{f,0}(\omega) \bm{\Sigma}_f(\omega) \bm{G}_f(\omega).
\end{equation}
If we can calculate the Green's function of the cluster $\bm{G}_f$, we obtain the self-energy via
\begin{equation}
\bm{\Sigma}_f(\omega) = \bm{G}_{f,0}^{-1}(\omega) - \bm{G}^{-1}_f(\omega).
\end{equation}
$\bm{G}_{f,0}$ is easily calculated, since it comes from a non-interacting system. Calculating $\bm{G}_f(\omega)$ amounts to solving the $N$, $N+1$ and $N-1$ many-body problems, which we now describe. 

\section{The Polynomial Expansion Algorithm}

To obtain $\bm{G}(\omega)$ we utilize the Chebyshev polynomial expansion method which we here describe briefly. For details see \cite{kpm}.

The Chebyshev polynomials of the first ($T_n$) and second ($U_n$) kind are defined recursively as
\begin{align} \label{eq:cherecur}
T_0(x) &= 1 & T_1(x) &= x  & T_{n+1}(x) &= 2xT_n(x) - T_{n-1}(x) \\
U_0(x) &= 1 & U_1(x) &= 2x & U_{n+1}(x) &= 2xU_n(x) - U_{n-1}(x).
\end{align}
They form orthogonal sets within the interval $x \in [-1, 1]$.

To simplify the equations we further define the functions $\phi_n(x) = \frac{T_n(x)}{\pi \sqrt{1 - x^2}}, $ which fulfill the orthogonality relation
\begin{align}
\la \phi_n | \phi_m \ra = \int_{-1}^1 \pi \sqrt{1 - x^2} \phi_n(x) \phi_m(x) dx = 
\frac{1 + \delta_{n0}}{2}\delta_{nm}.
\end{align}
We can expand any function $f(x)$ within $x \in [-1, 1]$ in terms of $\phi_n$ as
\begin{equation*} \label{eq:cheexp}
f(x) = \sum_{n = 0}^\infty g_n \frac{\la f | \phi_n \ra}{\la \phi_n | \phi_n \ra} \phi_n(x) =  \frac{\mu_0 g_0}{\pi \sqrt{1 - x^2}}  + 2\sum_{n = 1}^\infty g_n \mu_n \phi_n(x)  = \frac{1}{\pi \sqrt{1 - x^2}} \left[ g_0 \mu_0 + 2\sum_{n = 1}^\infty g_n \mu_n T_n(x) \right],
\end{equation*}
where the moments are given by $\mu_n = \int_{-1}^1 f(x) T_n(x) dx $. Since in practise we expand to a finite order $M$, we add additional factors $g_n$ to dampen oscillations in the resulting approximation. The problem of determining the optimal $g_n$ is described in detail in \cite{kpm}, and following their advice we choose $g_n = \frac{\sinh(4(1 - n/M)}{\sinh(4)}$. 

\subsection{Computing The Greens Function}

It turns out to be advantageous to first compute the spectral function $\bm{\mathcal{A}}(\omega)$, and obtain the corresponding Green's function from \cite{Stefanucci2013}
\begin{equation} \label{eq:G(A)}
\bm{G}(\omega) = \frac{1}{2\pi}\mathcal{P} \int_{-\infty}^\infty \frac{\bm{\mathcal{A}}(\omega')}{\omega - \omega'} d\omega' - \frac{i}{2}\bm{\mathcal{A}}(\omega).
\end{equation}
The $N$-particle spectral function can be obtained via the Lehmann representation \cite{Stefanucci2013} 
\begin{equation} \label{eq:spec}
\begin{split}
\mathcal{A}_{ij}(\omega) &= 2\pi\left( \sum_k \la \psi_0^N | \ao{c}{i} | \psi_k^{N + 1} \ra \la \psi_k^{N + 1} | \co{c}{j} | \psi_0^N \ra \delta(\omega - (E_k^{N+1} - E_0^N)) \right. \\
&\qquad + \left. \sum_l \la \psi_0^N | \co{c}{j} | \psi_l^{N - 1} \ra \la \psi_l^{N - 1} | \ao{c}{i} | \psi_0^N \ra \delta(\omega + (E_l^{N-1} - E_0^N)) \right)
\end{split}
\end{equation}
where $i$ and $j$ are compound quantum numbers containing both site and spin indices, and $\psi_k^N$ is the $k$:th eigenstate of the system with $N$ particles and energy $E_k^N$.

The moments are given by
\begin{align*}
\mu_n = \int_{-1}^1 \mathcal{A}_{ij}(\omega) T_n(\omega) d\omega 
= 2\pi\left( \sum_k \la \psi_0^N | \ao{c}{i} T_n(\hat{H} - E_0^N) | \psi_k^{N + 1} \ra \la \psi_k^{N + 1} | \co{c}{j} | \psi_0^N \ra \right. \\
+ \left. \sum_l \la \psi_0^N | \co{c}{j} T_n(E_0^N - \hat{H}) | \psi_l^{N - 1} \ra \la \psi_l^{N - 1} | \ao{c}{i} | \psi_0^N \ra \right),
\end{align*}
where substituting $\sum_n | \psi_n^{N \pm 1} \ra \la \psi_n^{N \pm 1} | = \mathbbm{1}$ and using $T_n(x) = (-1)^nT_n(-x)$ leads to
\begin{equation}
\mu_n = 2\pi \la \psi_0^N | \left( \ao{c}{i} T_n(\hat{H} - E_0^N) \co{c}{j} + (-1)^n \co{c}{j} T_n(\hat{H} - E_0^N) \ao{c}{i} \right) | \psi_0^N \ra.
\end{equation}
The spectral function is a sum of delta functions located at $\omega_n^{N+1} = E_n^{N+1} - E_0^N$ and $\omega_n^{N-1} = E_0^N - E_n^{N-1}$. Because the Chebyshev polynomials are orthogonal only within the interval $[-1,1]$, we need to shift and scale the energies to make sure that no poles fall outside the limits. Here we choose the zero of energy so that $E^N_0 = 0$ and furthermore scale the unit of energy by $\Delta > max(E_n^{N\pm 1}) + \varepsilon$, which guarantees that $\omega_n^{N\pm 1}/ \Delta = \bar{\omega}_n^{N\pm 1} \in [-1, 1]$ for all $n$. A small $\varepsilon >0 $ ensures that the spectrum does not reach the edges of the interval, where stability problems might arise. The rescaled moments become 
\begin{equation}
\bar{\mu}_n = 2\pi \la \psi_0^N | \left( \ao{c}{i} T_n(\hat{\bar{H}}) \co{c}{j} + (-1)^n \co{c}{j} T_n(\hat{\bar{H}}) \ao{c}{i} \right) | \psi_0^N \ra = \bar{\mu}_n^A + (-1)^n\bar{\mu}_n^R.
\end{equation}
where $\hat{\bar{H}} = (\hat{H} - E_0)/\Delta$, and we have defined the addition ($\bar{\mu}_n^A$) and removal ($\bar{\mu}_n^R$) moments.

Three Hamiltonian matrices $\bar{\bm{H}}^N, \bar{\bm{H}}^{N+1}$ and $\bar{\bm{H}}^{N-1}$ are needed for the computation. First the ground state $| \psi_0^N \ra$ is found through Lanczos iteration of $\bar{\bm{H}}^N$ \cite{Lanczos1950}. Then the moments for addition and removal processes are computed separately. For the addition part we first compute and store the states $| \alpha_i \ra = \co{c}{i} | \psi_0^N \ra$ and $| \alpha_j \ra = \co{c}{j} | \psi_0^N \ra$. Then, denoting $T_n(\bar{\bm{H}}^{N+1})| \alpha_j \ra \equiv | \alpha_j^n \ra$, the application of \Eq{eq:cherecur} gives us
\begin{equation}
\begin{split}
| \alpha_j^0 \ra &= T_0 (\bar{\bm{H}}^{N+1}) | \alpha_j \ra = |  \alpha_j \ra, \\
| \alpha_j^1 \ra &= T_1 (\bar{\bm{H}}^{N+1}) | \alpha_j \ra = \bar{\bm{H}}^{N+1} | \alpha_j^0 \ra, \\
| \alpha_j^{n+1} \ra &= 2 \bar{\bm{H}}^{N+1} | \alpha_j^n \ra - | \alpha_j^{n-1} \ra.
\end{split}
\end{equation}
This recursion represents the main numerical effort in the algorithm. It consists chiefly of matrix-vector multiplications (generally with a very sparse matrix), which is an operation that can be very efficiently parallelized.

After performing the iteration to a sufficient order, the moments are obtained by computing the inner product $ \bar{\mu}_n^A = 2\pi \la \alpha_i | \alpha_j^n \ra$. An analogous process gives the removal moments.

The resulting expansion
\begin{equation} \label{eq:specexp}
\bar{\mathcal{A}}_{ij}(\omega) \approx \frac{1}{\pi \sqrt{1 - \omega^2}} \left[ g_0 \bar{\mu}_0 + 2\sum_{n = 1}^N g_n \bar{\mu}_n T_n(\omega) \right]
\end{equation}
can now be inserted into \Eq{eq:G(A)} to obtain
\begin{equation} 
\bar{G}_{ij}(\omega) = -\frac{1}{\pi}\left( \frac{i g_0 \bar{\mu}_0}{2\sqrt{1 - \omega^2}} + \sum_{n = 1}^\infty g_n \bar{\mu}_n \left[U_{n-1}(\omega) + \frac{iT_n(\omega)}{\sqrt{1 - \omega^2}} \right] \right),
\end{equation}
where we have used the fact that $\bar{\mathcal{A}}_{ij}(\omega) = 0$ when $\omega \not\in [-1,1]$ and the identities \cite{Abramowitz}
\begin{equation}
\mathcal{P} \int_{-1}^1 \frac{1}{\sqrt{1 - y^2}(x - y)} dy = 0, \qquad \mathcal{P} \int_{-1}^1 \frac{T_n(y)}{\sqrt{1 - y^2}(x - y)} dy = -\pi U_{n-1}(x)
\end{equation}
to handle the integral.
Finally, we reverse the energy scaling by $G_{ij}(\omega) = \bar{G}_{ij}\left(\frac{\omega}{\Delta}\right)$.

\subsection{Finite Temperature Case}

As an aside, we mention that the Chebyshev method can also be utilized in the case of finite temperature. The spectral function is then given by \cite{Stefanucci2013}
\begin{equation} \label{eq:spec_temp}
\begin{split}
\mathcal{A}_{ij}(\omega; \beta) &= 2\pi \sum_q \rho_q \left( \sum_k \la \psi_q^N | \ao{c}{i} | \psi_k^{N + 1} \ra \la \psi_k^{N + 1} | \co{c}{j} | \psi_q^N \ra \delta(\omega - (E_k^{N+1} - E_q)) \right. \\
&\qquad\qquad + \left. \sum_l \la \psi_q^N | \co{c}{j} | \psi_l^{N - 1} \ra\la \psi_l^{N - 1} | \ao{c}{i} | \psi_q^N \ra \delta(\omega + (E_l^{N-1} - E_q^N)) \right),
\end{split}
\end{equation}
where $\rho_q = \frac{e^{-\beta E^N_q}}{Z}$, with $\beta = 1/{k_B T}$ the inverse temperature and $Z = \sum_n e^{-\beta E^N_n}$ the partition function.

By defining the functions \cite{kpm}
\begin{align}
J^P_{ij}(x, y) &= \sum_{q,k} \la \psi_q^N | \ao{c}{i} | \psi_k^{N + 1} \ra \la \psi_k^{N + 1} | \co{c}{j} | \psi_q^N \ra \delta(x - E_k^{N+1}) \delta(y - E_q^N), \\
J^H_{ij}(x, y) &= \sum_{q,l} \la \psi_q^N | \co{c}{j} | \psi_l^{N - 1} \ra\la \psi_l^{N - 1} | \ao{c}{i} | \psi_q^N \ra \delta(x - E_l^{N-1}) \delta(y - E_q^N)
\end{align}
we can express the spectral function as
\begin{equation}
\mathcal{A}_{ij}(\omega; \beta) = \frac{2\pi}{Z} \int dy \left[ J^P_{ij}(y + \omega, y) + J^H_{ij}(y - \omega, y) \right] e^{-\beta y},
\end{equation}
and expand it using a two dimensional Chebyshev expansion.

The moments for $\bar{J}^P_{ij}(x, y)$ and $\bar{J}^H_{ij}(x, y)$ (scaled so that $x,y \in [-1,1]$) are
\begin{equation}
\begin{split}
\bar{\mu}^P_{nm} &= \int_{-1}^1 \bar{J}^P_{ij}(x, y) T_n(x) T_m(y) dx dy 
= \sum_{q,k} \la \psi_q^N | \hat{c}_i T_n(\hat{\bar{H}}) | \psi_k^{N + 1} \ra \la \psi_k^{N + 1} | \hat{c}^\dagger_j T_m(\hat{\bar{H}}) | \psi_q^N \ra \\
&= Tr[\hat{c}_i T_n(\hat{\bar{H}}) \hat{c}^\dagger_j T_m(\hat{\bar{H}})], \\
\bar{\mu}^H_{nm} &= Tr[\hat{c}^\dagger_j T_n(\hat{\bar{H}}) \hat{c}_i T_m(\hat{\bar{H}})].
\end{split}
\end{equation}

The trace over the state space can be calculated efficiently by summing over a small random sample of states \cite{kpm}. One therefore does not need to solve any eigenstates. Note also that the main numerical effort is in calculating $J^P_{ij}(x, y)$ and $J^H_{ij}(x, y)$, which are independent of temperature. Storing these functions thus allows for efficient computation of quantities over temperature ranges.

\section{Application to a Finite Anderson Impurity}

As the simplest case of transport through an interacting system, we consider transport through a quantum dot, i.e. the single-impurity Anderson model. We consider one-dimensional leads, with the first few sites taken to be a part of the central cluster. By extending the cluster we hope to obtain a self-energy closer to that of the full embedded system. If the leads are identical, we can handle an arbitrary number of them by utilizing the rotation symmetry of the system. This we demonstrate below. In the same framework, we can also allow for hopping between adjacent leads.

The Hamiltonian for a quantum dot connected to $N_l$ one-dimensional leads ($n$) with $N_r$ sites ($i$, $j$) in each is
\begin{equation} \begin{split}
\hat{H} = &-\sum_n \sum_{i, j, \sigma} t_{ij} \cs{c}{n,i} \as{c}{n,j} - \sum_n \sum_\sigma t_0 \left( \cs{d}{} \as{c}{n,1} + \cs{c}{n,1} \as{d}{} \right) \\
&- \sum_{\la n, m \ra} \sum_{i, \sigma} \tau_i \cs{c}{n,i} \as{c}{m,i} + V (\au{n}{d} + \ad{n}{d}) + U \au{n}{d} \ad{n}{d},
\end{split} \end{equation}
where the first term contains hopping along the leads, second is the lead-dot connection, third connects neighboring leads, and fourth and fifth are the dot potential and interaction.

\begin{center}
\begin{figure} 
\begin{equation}
\begin{array}{ccc} 
\vcenter{\hbox{\includegraphics[width=0.24\textwidth]{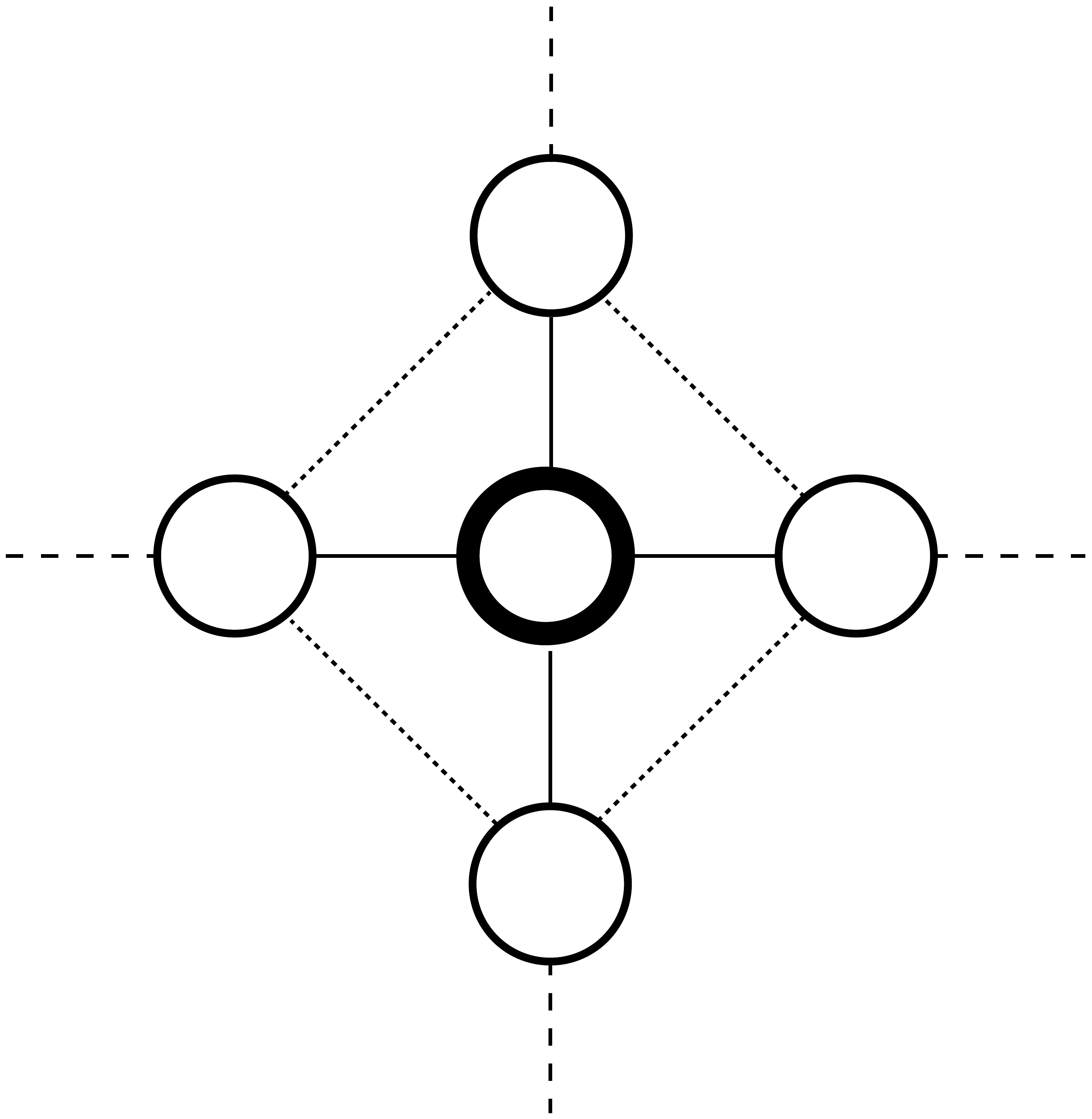}}} & \Scale[3]{\Rightarrow} & \vcenter{\hbox{\includegraphics[width=0.15\textwidth]{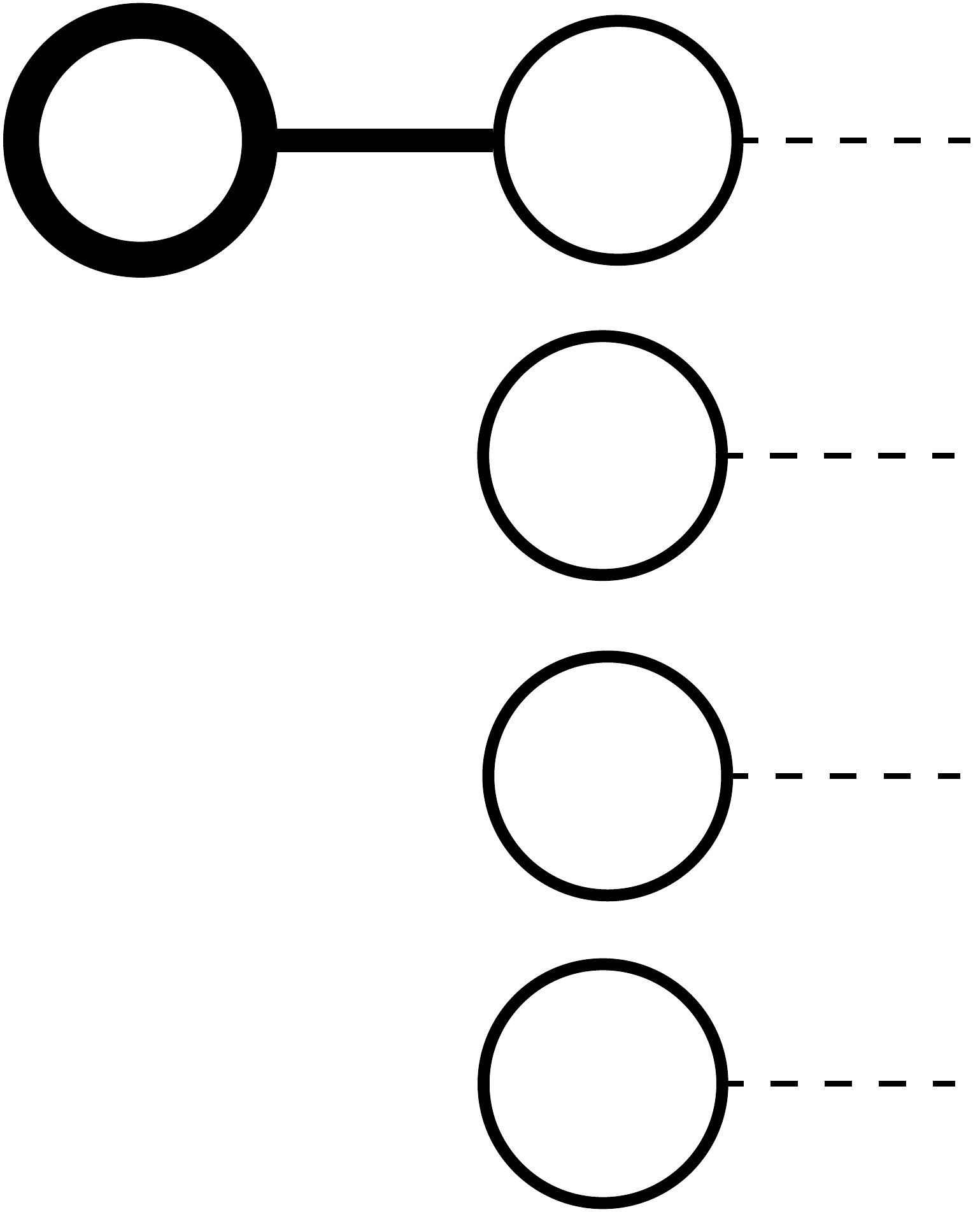}}}
\end{array}
\end{equation}
\caption{A graphical representation of the unitary transformation, \Eq{eq:transformation}, for a system with four leads.}\label{fig:transformation}
\end{figure}
\end{center}

We exploit the symmetry of this system by a unitary transformation, defining the operators
\begin{equation} \label{eq:transformation}
\as{a}{k,i} = \frac{1}{\sqrt{N_l}} \sum_{n=1}^{N_l} e^{i\frac{2\pi}{N_l}(n-1)(k-1)} \as{c}{n,i},
\end{equation}
in terms of which the Hamiltonian takes the form
\begin{equation} \begin{split} \label{eq:transformedH}
\hat{H} &= \sum_{k=1}^{N_l} \hat{H}_k \\ 
&= -\sum_{i, j, \sigma} t_{ij} \cs{a}{1,i} \as{a}{1,j} - \sum_\sigma t'_0 \left( \cs{d}{} \as{a}{1,1} + \cs{a}{1,1} \as{d}{} \right) - W_1 \sum_i \tau_i \ao{n}{1,i}
+ V (\au{n}{d} + \ad{n}{d}) + U \au{n}{d} \ad{n}{d} \\
&- \sum_{k=2}^{N_l} \left[ \sum_{i, j, \sigma} t_{ij} \cs{a}{k,i} \as{a}{k,j} + W_k \sum_i \tau_i \ao{n}{k,i} \right],
\end{split} \end{equation}
where $W_k = 2 \cos\left(\frac{2\pi}{N_l}(k-1)\right)$ and $t'_0 = \sqrt{N_l}t_0$. The Hamiltonian is therefore divided into $N_l$ independent commuting terms corresponding to different wavenumbers for rotation of phase when going around the dot. Only the $k = 1$ part is coupled to the dot, the remaining $N_l - 1$ subsystems are non-interacting. See figure \ref{fig:transformation} for an illustration of this transformation.

The spectral function of the whole system $\mathcal{A}$ can be expressed in terms of the spectral functions $\mathcal{A}_k$ of the $N_l$ subsystems:
\begin{equation} \label{eq:spec_decomp}
\mathcal{A}_{ij}(\omega) =
\left\lbrace \begin{array}{ll}
\mathcal{A}_{1,ij}(\omega) & \text{when $i$ and $j$ are both the dot site,} \\
\frac{1}{\sqrt{N_l}} \mathcal{A}_{1,ij}(\omega)  & \text{when either $i$ or $j$ is the dot site,} \\
\frac{1}{N_l} \sum_k e^{-i\frac{2\pi}{N_l}(k-1)(n - m)} \mathcal{A}_{k,ij}(\omega) & \text{when $i$ and $j$ are in the leads $n$ and $m$ respectively,}
\end{array} \right.
\end{equation}
where $\mathcal{A}_1$ can be calculated using the method described in the previous chapter, and $\mathcal{A}_k$ are easily obtained, being spectral functions of non-interacting systems.

\section{Results for the Finite Anderson Impurity}

\begin{figure} 
\begin{minipage}[]{0.5\textwidth}
\subfloat[\textcolor{red}{Spectral function} and \textcolor{blue}{self-energy} in the dot for a varying lead length $N_r$ in the case of two leads.]{\includegraphics[width=\textwidth,height=0.75\textwidth  \label{fig:disconnected_resultsa}
]{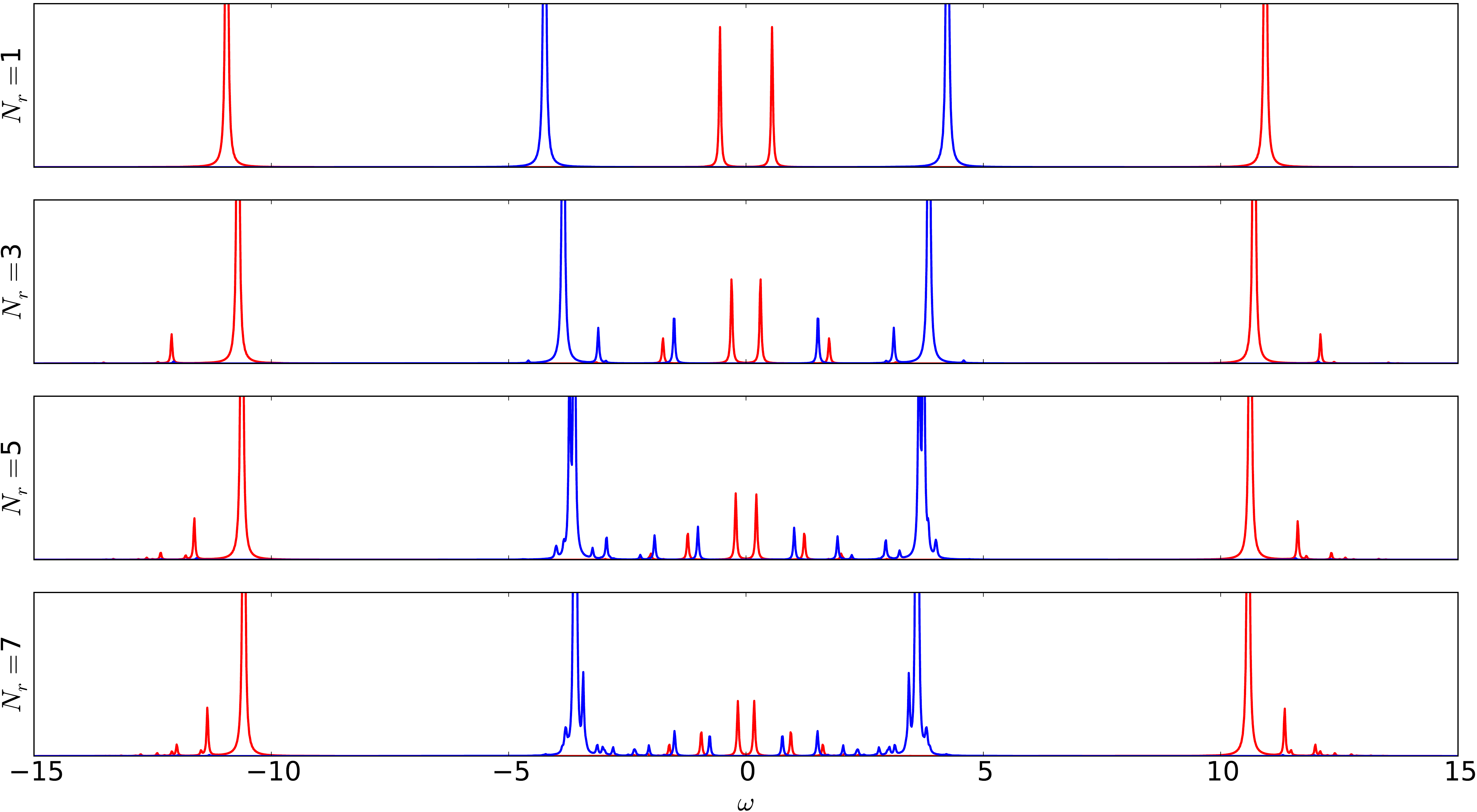}}
\vfill
\subfloat[\textcolor{red}{Spectral function} and \textcolor{blue}{self-energy} in the dot for a varying lead hopping $t$ in the case of two three site leads. Hopping to the dot is kept constant at $t_0 = 1$]{\includegraphics[width=\textwidth,height=0.75\textwidth
 \label{fig:disconnected_resultsb}
 ]{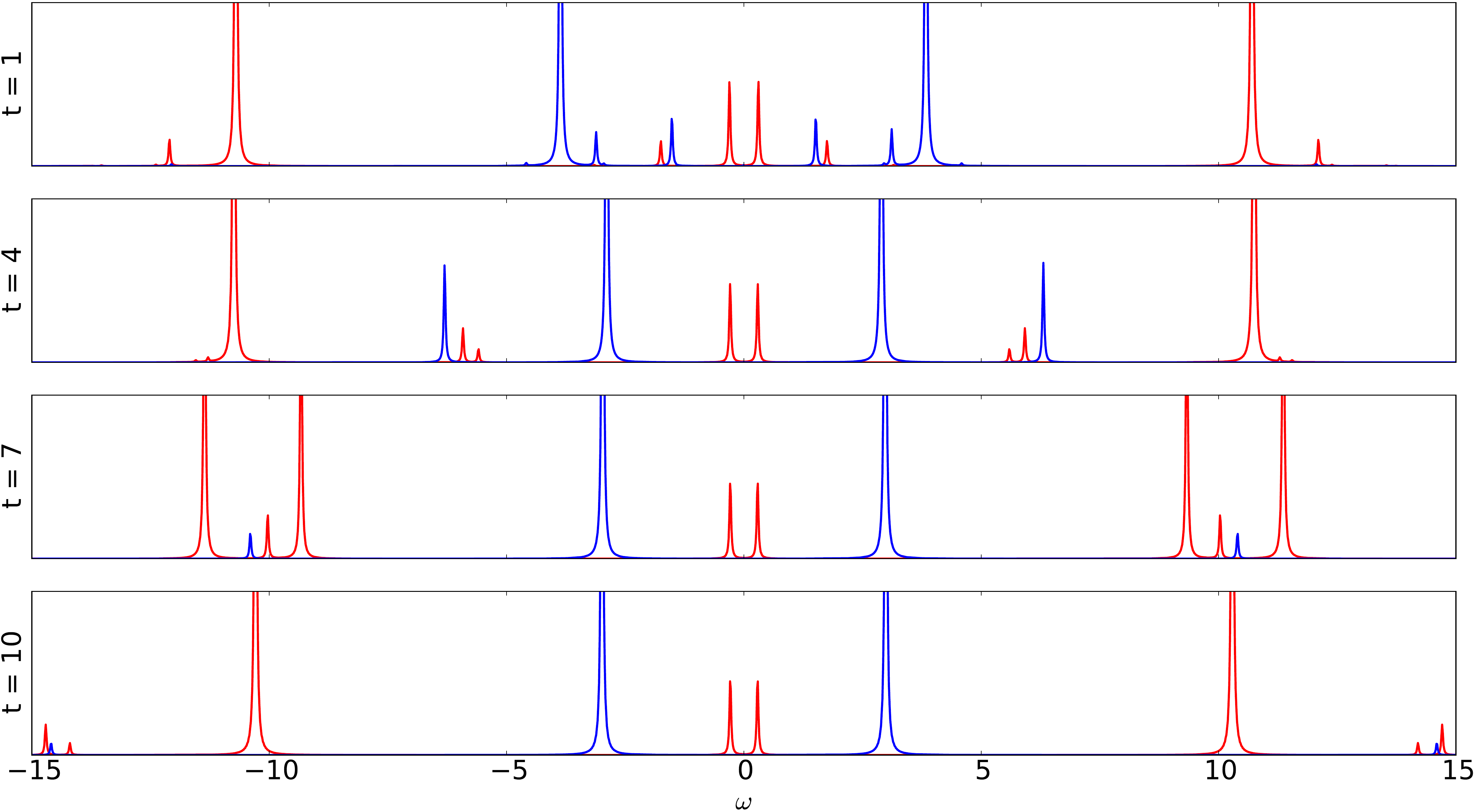}}
\end{minipage}
\begin{minipage}[]{0.5\textwidth}
\subfloat[\textcolor{red}{Spectral function} and \textcolor{blue}{self-energy} in the dot for a varying dot potential $V$ in the case of two single site leads.
 \label{fig:disconnected_resultsc}
 ]{\includegraphics[width=\textwidth,height=1.66\textwidth]{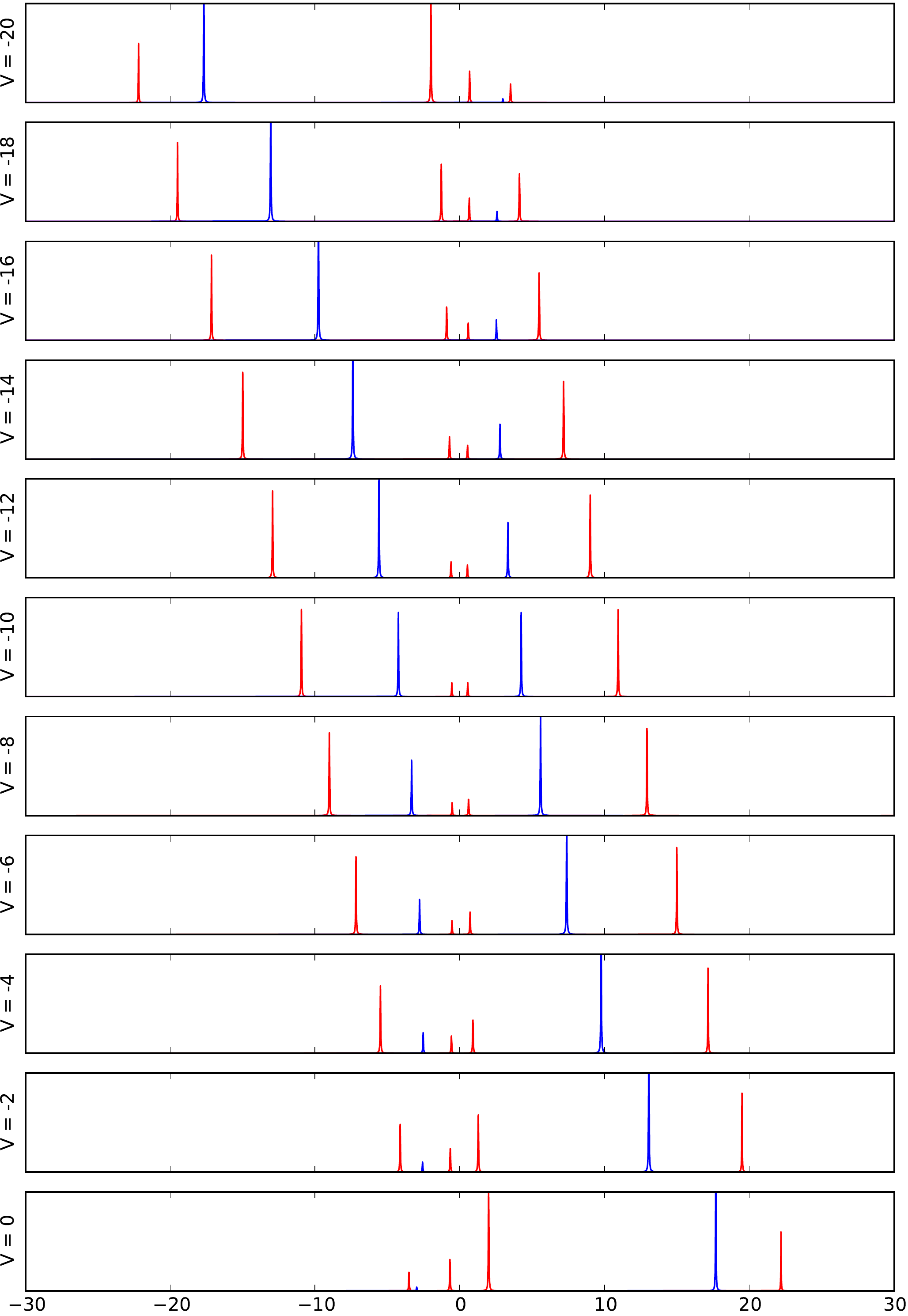}}
\end{minipage}%
\caption{}\label{fig:disconnected_results}
\end{figure}

Figure \ref{fig:disconnected_results} shows results obtained using the above machinery for finite Anderson systems. We show the dot-to-dot components of $\mathcal{A}(\omega)$ and $-\Im m \Sigma (\omega)$, using interaction strength $U = 20$, dot potential $V = -10$ and nearest neighbor hopping $t = 1$ unless otherwise stated. Hopping between leads is here turned off ($\tau_i = 0$). In all cases the system is spin symmetric with an equal number of spin up and spin down particles, and either at half-filling or one particle over half-filling (if the number of sites is odd).

In figure \ref{fig:disconnected_resultsa} we increase the lead length in a system with two leads. The spectral functions here have a distinct structure: a central peak cluster around $\omega = 0$, and two side peak clusters on either side. One can understand the structure in terms of lead/dot occupation. The side clusters contain transitions where the dot occupation is changed (addition on the right, removal on the left), while the central cluster contains transitions where the dot occupation remains relatively unchanged and instead the lead occupation changes. Each cluster corresponds to a different dot occupation, and therefore the energy scale separating them is determined by the dot potential and interaction. The peaks within each cluster, on the other hand, correspond to lead excitations, and the scale of their spread is determined by the scale of the hopping terms in the leads. This is a helpful picture, but note that it is only approximate, and predicated on the fact that the two energy scales are here distinct.

From figure \ref{fig:disconnected_resultsa} we see that increasing the lead length results in additional peaks appearing in each cluster. These can be understood to represent new possible transitions corresponding to new lead states. 

Figure \ref{fig:disconnected_resultsb} shows the effect of increasing the hopping along the leads (with dot-to-lead hopping kept constant). This increases the energy related to lead excitations causing the peak clusters to widen. When $t = 7$ one of the lead state transitions hits a resonance with a dot state transition, leading to hybridization of the corresponding states, and the side peaks splitting as a result. This signals the break down of the simplified picture represented above, since the hybridized states are no longer associated with a specific dot occupation value.

It is tempting to imagine the limit of infinitely long leads and high lead hopping, which corresponds to the Anderson impurity model in the Kondo regime. The spectral function in this limit contains broadened Coulomb blockade side peaks at $\omega = \pm10$, as well as a sharp Kondo peak at $\omega = 0$ \cite{CuevasBook}. The results in figure \ref{fig:disconnected_resultsa} show all the signs of approaching the correct limit, with the side peaks moving toward $\omega = \pm10$ and the central peaks inching toward each other to, presumably, merge at $\omega = 0$. Furthermore the resonance between lead transitions and dot transitions would in the limit of infinite number of side peaks in the central cluster lead to the broadening of the side peaks, while the absence of similar process for the central peaks might conceivably leave them sharp.

In figure \ref{fig:disconnected_resultsc} the dot potential $V$ is shifted. Note that the central peak cluster shows resistance to the change in $V$, while the side peak clusters shift roughly linearly with $V$. This is analogous to the situation in transport in the Kondo regime \cite{CuevasBook}, where the Kondo peak is resistant to the gate potential, while the Coulomb blockade side peaks will shift linearly.

We note that the functions are even functions of $\omega$. The spectral functions, \Eq{eq:spec}, consisting of particle addition and removal parts, will be symmetric if the ground state of the system is particle-hole symmetric, i.e. if the system is at half-filling and the occupation is uniform. A half-filled state will have uniform occupation if the interaction strengths and site potentials are chosen so that the energy costs related to adding or removing a particle from a singly occupied site are equal. This corresponds to $V_i = -\frac{U_i}{2}$ in the systems considered here. Figure \ref{fig:disconnected_resultsc} demonstrates how the symmetry is broken when the potential is shifted.

We see, however, that spectral functions in figure \ref{fig:disconnected_resultsa} are symmetric even though the system as a whole is not at half-filling (having an odd number of sites). This behavior becomes more clear after the symmetry transformation, when the interacting subsystem is left with an even number of sites. If lead-to-lead hopping is not allowed, our calculations have always shown this subsystem to be at half-filling on the ground state, which explains the symmetry. The non-interacting subsystems may not be at half-filling, but this does not matter since they do not contribute to the dot-to-dot component of $\mathcal{A}$, as seen from \Eq{eq:spec_decomp}.

Allowing inter-lead hopping can again break the symmetry. This is because it causes energy shifts in the subsystems (the $W_k$ terms in \Eq{eq:transformedH}). The energy shifts can lead to a different particle distribution among the subsystems. Because of this the interacting subsystem ($k=1$) can move away from half-filling, breaking the particle-hole symmetry and thus the symmetry of the spectral function.

\section{Implantation Method}
Using the techniques described in the previous sections, we can obtain the \emph{exact} interaction self-energy for a finite system. In general, this procedure will yield a self-energy matrix, which can then be implanted into Eq. (\ref{eq:dyson}) to obtain the embedded Green's function. In order to simplify the discussion, however, we will continue with the single-impurity Anderson system. In this case only the dot-to-dot component of the interaction self-energy is non-zero, and Eq. (\ref{eq:dyson}) becomes a scalar equation for the dot-to-dot component of the retarded Green's function
\begin{align} \label{eq:infinite_greens_function}
G(\omega) = \frac{1}{\omega - V - \Sigma (\omega) - \Sigma_{em}(\omega) }.
\end{align}
For an Anderson impurity connected to finite leads, we have 
\begin{align}
G_f(\omega) = \frac{1}{\omega - V - \Sigma_{f} (\omega) - \Sigma_{f,em}(\omega) },
\end{align}
where $f$ stands for finite. The embedding self-energy now results from a finite number of sites, and the interaction self-energy only describes interaction events of the finite system. 

We form an approximate $G$ for the infinite system by implanting $\Sigma_{f}$ into \Eq{eq:infinite_greens_function}
\begin{align}\label{eqAnderson}
G(\omega) \approx \frac{1}{\omega - V - \Sigma_{f} (\omega) - \Sigma_{em}(\omega) },
\end{align}
where the dot is now embedded to infinite leads, but the interaction events are contained within the parts of the leads included in our finite system. Increasing the size of the finite system will improve the approximation, ultimately reaching the exact result in the limit. A natural question is then, how big is big enough? 

\section{Results for the Implantation Method}

\begin{figure}
 \centering
 \includegraphics[width=0.7\textwidth]{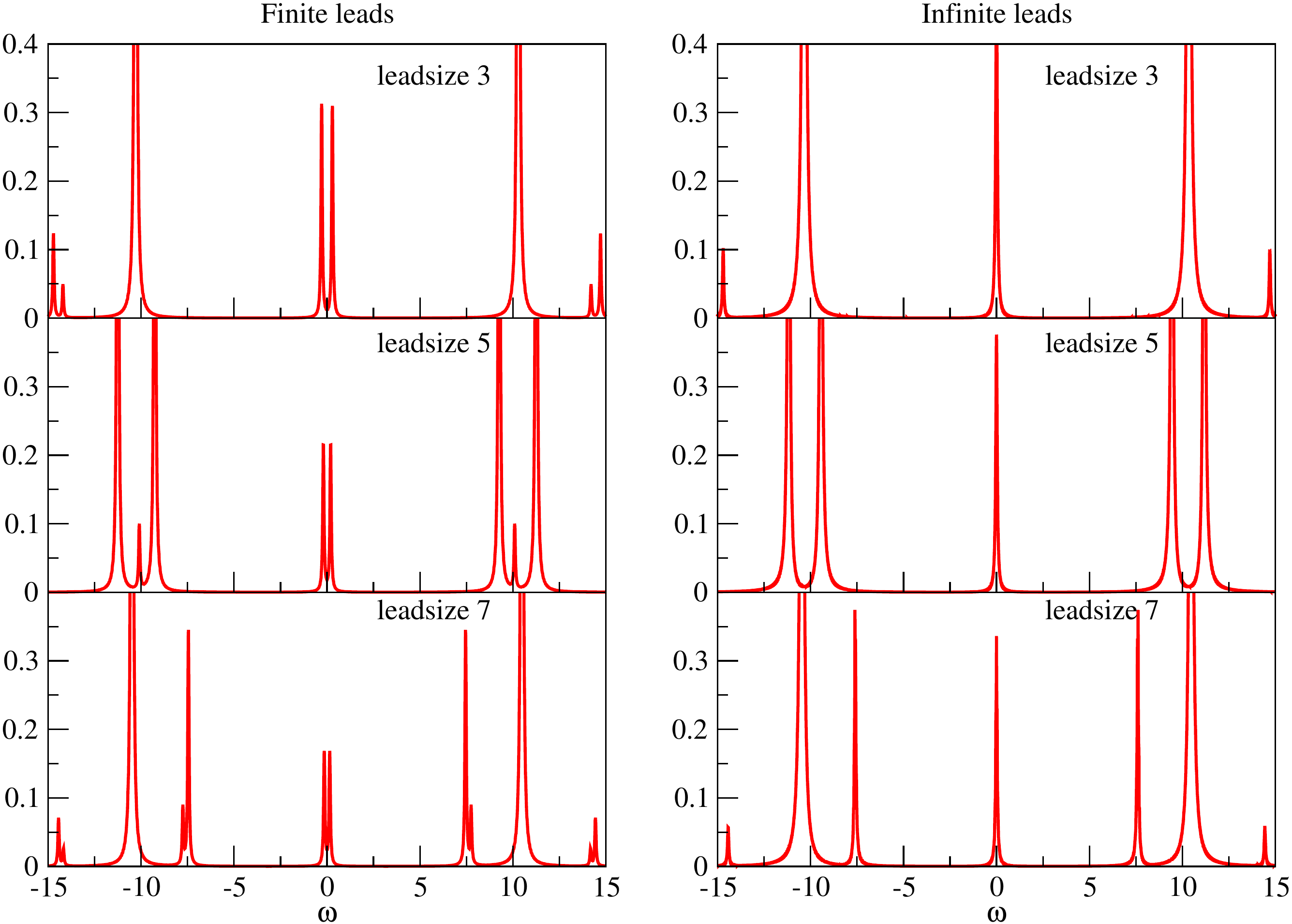}
 \caption{ {\color{red}Spectral function} on the impurity site, for different numbers of sites in two identical leads. Left: The exact $A$ from the finite system. Right: $A$ using the implantation method. }
 \label{figImplantation}
\end{figure}

In order to see how the implantation method fares, we solve the finite lead-dot-lead system exactly and construct the exact interaction self-energy $\Sigma_f$. We perform calculations for different lead lengths in order to see how the results change. Due to symmetry considerations, the leads have to be identical, and thus we have an odd number of sites in our system. We consider $N_\uparrow = N_\downarrow = (N_s+1)/2$, and put $V = -U/2=-10$ at the central site. We then calculate $G_f(\omega)$ and $\Sigma_{f,int}(\omega)$ using the polynomial expansion method, and then implant $\Sigma_{f,int}$ into the single-impurity Anderson model via \Eq{eqAnderson}. Here, we use semi-infinite tight-binding leads\cite{Stefanucci2013}, and $\mu = 0$. For different fillings, we can define $\mu$ by forcing the density of the impurity site in the infinite system to coincide with the one from the finite system. In this section, we use a different normalization of $A = \mathcal{A} / (2\pi)$, to be more consistent with the literature. 

The spectral functions for the finite and infinite Anderson models are shown in Figure \ref{figImplantation}. The spectral functions for the finite systems have small but finite peak widths, which is due to them being expressed in terms of polynomials of finite order \cite{kpm}. We note that there is no peak at $A(0)$. In the implantation scheme, the spectral features are broadened in an inhomogeneous way. Furthermore, we obtain a peak at $\omega = 0$. Apart from the additional sharp peaks, we obtain the characteristic three-peak structure\cite{CuevasBook} with two Coulomb blockade peaks at $\omega = \pm U/2$ and a sharp Kondo peak at $\omega = 0$. The features broadly agree with the true spectral function of the Anderson model in this parameter regime.

In order to understand the broadening, we study in more detail the spectral function. In the formulas, we treat the semi-infinite leads in the wide-band limit approximation, which means that we take the embedding self-energy to be purely imaginary and equal to its value at $\mu,$ $\Sigma_{em}(0) = i\Gamma$, where $\Gamma = 2 t_0^2 / t = 0.2$. This will slightly overestimate the broadening of the Coulomb peaks. We then get from \Eq{eqAnderson}
\begin{align}
 A(\omega) = \frac{1}{\pi} \frac{\Gamma - \Im m \Sigma_f (\omega)}{ (\omega - V - \Re e \Sigma_f (\omega) )^2 + (\Gamma - \Im m \Sigma_f (\omega) )^2 }.
\end{align}
For our finite system $V \approx -\Re e \Sigma_f (0)$, which means that we obtain a peak at $\omega = 0$ of height $\frac{1}{\pi (\Gamma - \Im m \Sigma_f (0) )}$. 
For well-behaved systems $\Im m \Sigma_f (\mu) = 0$ \cite{Stefanucci2013}, which is true for the exact solution of the Anderson model. In our numerics, however, the polynomial expansion broadens the self-energy, and thus our Kondo peak height is smaller than the correct value of $A(0) = \frac{1}{ \pi \Gamma}$\cite{Langreth1966}.

We now turn to the inhomogeneous broadening. 
The Coulomb blockade peaks are broadened by the order of $\Gamma$, while the Kondo peak is broadened less. This is due to the frequency dependence of $\Re e \Sigma_f (\omega)$ ($ \Sigma_{em}(\omega)$ has a small $\omega-$dependence in this limit). By assuming that $\Im m \Sigma_f (\omega)$ depends weakly on $\omega$ at the peak positions $\omega_0$ ($\Im m \Sigma_f (\omega_0)$ can be absorbed into $\Gamma$ without changing the conclusions, also see Figure \ref{fig:disconnected_resultsb}), 
the Full-Width at Half-Maximum of the peaks is $ \text{FWHM} =  2\frac{\Gamma}{1-\frac{\partial }{\partial \omega} \Re e \Sigma_f (\omega_0)  }$.

At the Coulomb blockade peaks, $\Sigma_f$ depend weakly on $\omega$, giving FWHM $\sim \Gamma$. Around $\omega = 0$, $-\Re e \Sigma'_f$ is much larger, reducing the FWHM. However, compared to the true width of the Kondo peak, exponentially supressed as a function of $U$ \cite{Wilson1975}, the implantation method yields a peak much too wide. In order to improve the description, $\Sigma_f$ should have a more negative slope at $\omega = 0$. We have indeed seen that the slope increases as a function of cluster size, hinting that the implantation method has the possibility of giving a sharp Kondo peak. 

Future studies will be done to reduce the artifical broadening from the polynomial expansion. We have seen that (somewhat artificially) using the same $\Sigma_f$ as before but using $\Gamma = 2$ broadens the Coulomb peaks with $\sim \Gamma$ (much bigger than the artifical smoothening), but does not affect the Kondo peak width much. 

\section{Conclusions}
In this work, we have given an overview of the polynomial expansion algorithm and applied it to obtain spectral functions for single-impurity Anderson systems with a high degree of symmetry. By making use of a unitary transformation, the original system disconnects into a smaller interacting one, as well as several non-interacting chains. In systems fulfilling particle-hole symmetry, the spectral function is symmetric ($\mathcal{A}(\omega) = \mathcal{A}(-\omega)$), which simply means that the addition and removal parts of $\mathcal{A}$ are identical. However, we found systems which are not particle-hole symmetric, yet still have a symmetric spectral function. This we explained by noting that in these cases, the reduced interacting cluster was particle-hole symmetric, showing a hidden symmetry for the spectral function. 

By calculating the self-energy from an isolated system and implanting it into a transport setup, we obtained spectral functions for the Anderson model. This can be seen as a perturbation expansion from an isolated system, but on the self-energy level. The implantation of an exact self-energy from a finite system yields a complex frequency dependent self-energy for the infinite system, which gives rise to highly non-trivial features in the spectral function, such as the Coulomb peaks and the Kondo peak. The frequency dependence of the self-energy led to an inhomogeneous broadening of the spectral peaks. The Kondo peak was too broad compared to the exact solution, which we attribute to using a too small cluster in the calculation of the self-energy. We have seen that increasing the lead sizes gives a sharper Kondo peak. 

\section*{Acknowledgements}
M.H. would like to thank the Väisälä foundation for support.
D.K. and R.v.L. would like to thank the Academy of Finland for support. We thank Gianluca Stefanucci for useful discussions.


\section*{References}
\begin{thebibliography}{9}

\bibitem{Heidrich-Meisner2009}
F. Heidrich-Meisner, 
G. B. Martins,
C. A. Büsser,
K. A. Al-Hassanieh,
A. E. Feiguin,
G. Chiappe,
E. V. Anda and 
E. Dagotto,
Eur. Phys. J. B {\bf 67}, 527-542 (2009)

\bibitem{Jacob2010}
D. Jacob, K. Haule and G. Kotliar,
Phys. Rev. B {\bf 82}, 195115 (2010)

\bibitem{Arrigoni2013}
E. Arrigoni, M. Knap and W. von der Linden,
Phys. Rev. Lett. \textbf{110}, 086403 (2013)

\bibitem{Ishida2012}
H. Ishida and A. Liebsch,
Phys. Rev. B \textbf{86}, 205115 (2012)

\bibitem{Ferrari1999}
V. Ferrari, G. Chiappe, E. V. Anda and Maria A. Davidovich,
Phys. Rev. Lett. {\bf 82}, 5088 (1999)

\bibitem{Hubbard1963}
J. Hubbard, 
Proc. R. Soc. London A {\bf 276}, 238 (1963)

\bibitem{Gros1993}
C. Gros and R. Valent\'{\i},
Phys. Rev. {\bf 48} 418 (1993) 	
	
\bibitem{Potthoff2003}
M. Potthoff,
Eur. Phys. J. B \textbf{32}, 429-436 (2003)

\bibitem{kpm}
A. Weibe, G. Wellein, A. Alvermann and H. Fehske,
Rev. Mod. Phys., \textbf{78}, 275 (2006)

\bibitem{Stefanucci2013}
G. Stefanucci and R. van Leeuwen,
\emph{Nonequilibrium Many-Body Theory of Quantum Systems: A Modern Introduction},
Cambridge University Press (2013)

\bibitem{Wilson1975}
K. G. Wilson,
Rev. Mod. Phys. {\bf 47} 773 (1975)

\bibitem{Hershfield1991}
S. Hershfield, J. H. Davies and J. W. Wilkins, 
Phys. Rev. Lett. {\bf 67}, 3720 (1991)

\bibitem{Langreth1966}
D. C. Langreth, 
Physical Review {\bf 150}, 516 (1966).

\bibitem{CuevasBook}
J. C. Cuevas and E. Scheer,
\emph{Molecular Electronics: An Introduction to Theory and Experiment}, 
World Scientific Publishing (2010)

\bibitem{Abramowitz}
M. Abramowitz and I. A. Stegun (Eds.)
\emph{Handbook of Mathematical Functions with Formulas, Graphs, and Mathematical Tables},
Dover, New York (1970)

\bibitem{Lanczos1950}
C. Lanczos,
J. Res. Natl. Bur. Stand. {\bf 45} 255 (1950)

\end{thebibliography}
\end{document}